\documentstyle[12pt]{article}
\title{Tensor susceptibility of the QCD vacuum from an effective 
       quark-quark interaction}
\author{Hongting Yang $^a$, Hongshi Zong $^a$, 
        Jialun Ping $^b$, Fan Wang $^a$ \\
       {\small\sl $^a$ Department of Physics, Nanjing University, 
       Nanjing 210093, P.R. China}\\
       {\small\sl $^b$ Department of Physics, Nanjing Normal University, 
       Nanjing 210097, P.R. China}} 
\date{}

\begin{document}
\maketitle
\begin{abstract}
Treating the bilocal quark-quark interaction kernel as an input parameter, the
self-energy functions can be determined from the ``rainbow'' Dyson-Schwinger
equation, which is obtained in the global color symmetry model. The tensor 
susceptibility of QCD vacuum can be calculated directly from these self-energy
functions. The values we obtained are much smaller than the estimations from
QCD sum rules and from chiral constituent quark model.
\end{abstract}
\newpage

The tensor susceptibility of QCD vacuum is relevant for the determination of
nucleon tensor charge~(\cite{HX95},\cite{XJ97}), which is related to the first
moment of the transversity distribution $h_1(x)$~\cite{RX91}, where $h_1(x)$
is chiral-odd spin-dependent structure function and can be measured in the
polarized Drell-Yan process~\cite{JD79}. The previous estimations for the
value of tensor susceptibility were obtained by QCD sum rules
techniques~(\cite{HX96}-\cite{AS00}) or from chiral constituent quark
model~\cite{WM98}. In this letter, we report the different results of tensor
susceptibility from the global color symmetry model (GCM)
approximation~(\cite{RC85}-\cite{T97}) to QCD.

GCM based upon an effective quark-quark interaction can be defined through a
truncation of QCD as follows. The QCD partition function for massless quarks
in Euclidean space can be written as
\begin{equation}
 {\cal Z}=\int{\cal D}q{\cal D}\bar{q}e^{-\int{\rm d}x\bar{q}{\not\partial}q}
          e^{W[J]}
\end{equation}
with $W[J]$ given by
\begin{equation}
  e^{W[J]}=\int{\cal D}Ae^{\int{\rm d}x(-{1\over
  4}G^a_{\mu\nu}G^a_{\mu\nu}+J^a_{\mu}A^a_{\mu})},
\end{equation}
where $J^a_{\mu}(x)=ig\bar{q}(x)\gamma_{\mu}\frac{\lambda^a}{2}q(x)$.
The functional $W[J]$ can be formally expanded in terms of the current 
$J^a_{\mu}$:
\begin{equation}
W[J]={1\over 2}\int{\rm d}x{\rm d}yJ^a_{\mu}(x)D^{ab}_{\mu\nu}(x,y)J^b_{\nu}(y)
+\frac{1}{3!}\int J^a_{\mu}J^b_{\nu}J^c_{\rho}D^{abc}_{\mu\nu\rho}+\cdots.
\end{equation}
The GCM is defined through the truncation of the functional $W[J]$ in which
the higher order $n(\ge3)$-point functions are neglected, and only the gluon
2-point function $D^{ab}_{\mu\nu}(x,y)$ is retained. This is an effective
model based on the bilocal quark-quark interaction
$D^{ab}_{\mu\nu}(x,y)=D^{ab}_{\mu\nu}(x-y)$. This model maintains global color
symmetry of QCD. The primary loss by this truncation is local SU(3) gauge 
invariance.

By the functional integration approach, the partition function of this 
truncation can be given by
\begin{equation}     \label{pfGCM}
  {\cal Z}_{\rm GCM}=\int{\cal D}q{\cal D}\bar{q}\exp\left (
  -\int{\rm d}x\bar{q}{\not\partial}q-
 \frac{g^2}{2}\int{\rm d}x{\rm d}yj^a_{\mu}(x)  
 D^{ab}_{\mu\nu}(x-y)j^b_{\nu}(y)\right ),   
\end{equation} 
where $j^a_{\mu}(x)=\bar{q}(x)\gamma_{\mu}\frac{\lambda^a}{2}q(x)$ is the quark
color current. In~\cite{T97}, the gluon 2-point function
$D^{ab}_{\mu\nu}(x-y)$ is treated as the model input parameter, which is
chosen to reproduce the pion decay constant in the chiral limit $f_\pi=87$~MeV
and moreover reproduce values for the chiral low energy coefficients. For
simplicity we use a Feynman-like gauge
$D^{ab}_{\mu\nu}(x-y)=\delta_{\mu\nu}\delta^{ab}D(x-y)$. By the standard
bosonization procedure, the resulting expression for the partition function in
terms of the bilocal field integration is ${\cal Z}_{\rm GCM}=\int{\cal
D}{\cal B}e^{-S[{\cal B}]}$, where the action is given by 
\begin{equation}
 S[{\cal B}]=-{\rm TrLn}[G^{-1}]+\int{\rm d}x{\rm d}y
  \frac{{\cal B}^{\theta}(x,y){\cal B}^{\theta}(y,x)}{2g^2D(x-y)},
\end{equation}
and the quark inverse Green's function $G^{-1}$ is defined as
\begin{equation}
 G^{-1}(x,y)={\not\partial}\delta(x-y)+\Lambda^{\theta}{\cal B}^{\theta}(x,y).
\end{equation}
Here the quantity $\Lambda^{\theta}$ arises from Fierz reordering of the
current-current interaction term in~(\ref{pfGCM})
\begin{equation}
 \Lambda^{\theta}_{ji}\Lambda^{\theta}_{lk}=
 (\gamma_{\mu}\frac{\lambda^a}{2})_{jk}
 (\gamma_{\mu}\frac{\lambda^a}{2})_{li} 
\end{equation}
and is the direct product of Dirac, flavor SU(3) and color matrices:
\begin{equation}
  \Lambda^{\theta}={1\over 2}(1_D,i\gamma_5,\frac{i}{\sqrt{2}}\gamma_{\mu},
  \frac{i}{\sqrt{2}}\gamma_{\mu}\gamma_5)\otimes(\frac{1}{\sqrt{3}}1_F,
  \frac{1}{\sqrt{2}}\lambda^a_F)\otimes({4\over 3}1_c,
  \frac{i}{\sqrt{3}}\lambda^a_c). 
\end{equation}

The vacuum configurations are defined by minimizing the bilocal action:
$\left. \frac{\delta S[{\cal B}]}{\delta {\cal B}} \right |_{{\cal B}_0}=0,$
which gives 
\begin{equation}
  {\cal B}^{\theta}_0(x-y)=g^2D(x-y){\rm tr}[\Lambda^{\theta}G_0(x-y)].
\end{equation}
These configurations provide self-energy dressing of the quarks through the
definition
$\Sigma(p)\equiv\Lambda^{\theta}{\cal B}^{\theta}_0(p)=i{\not p}[A(p^2)-1]
+B(p^2)$. The self-energy functions $A$ and $B$ satisfy the so-called 
``rainbow'' Dyson-Schwinger equation,
\begin{eqnarray}         \label{RDSE}
   [A(p^2)-1]p^2 &=& {8\over 3}\int\frac{{\rm d}^4q}{(2\pi)^4}g^2D(p-q)
   \frac{A(q^2)q\cdot p}{q^2A^2(q^2)+B^2(q^2)}    \nonumber \\
          B(p^2) &=& {16\over 3}\int\frac{{\rm d}^4q}{(2\pi)^4}g^2D(p-q)
   \frac{B(q^2)}{q^2A^2(q^2)+B^2(q^2)}. 
\end{eqnarray}

In terms of $A$ and $B$, the quark Green's function at ${\cal B}^{\theta}_0$ 
is given by 
\begin{equation}
  G_0(x,y)=G_0(x-y)=\int\frac{{\rm d}^4p}{(2\pi)^4}
  \frac{-i{\not p}A(p^2)+B(p^2)}{p^2A^2(p^2)+B^2(p^2)}e^{ip\cdot(x-y)}.
\end{equation} 
The vacuum expectation value of any operator of the form
\begin{equation}
  Q_n\equiv(\bar{q}_{j_1}\Lambda^{(1)}_{j_1i_1}q_{i_1})
  (\bar{q}_{j_2}\Lambda^{(2)}_{j_2i_2}q_{i_2})\cdots
  (\bar{q}_{j_n}\Lambda^{(n)}_{j_ni_n}q_{i_n})
\end{equation}
is
\begin{equation}     \label{Wick}
  \langle Q_n\rangle=(-1)^n\sum_p(-1)^p\left \{\Lambda^{(1)}_{j_1i_1}\cdots
   \Lambda^{(n)}_{j_ni_n}(G_0)_{i_1j_{p1}}\cdots(G_0)_{i_nj_{pn}}\right \},
\end{equation}
where $\Lambda^{(i)}$ represents an operator in Dirac, flavor and color 
space and $p$ stands for a permutation of $n$ indices~(\cite{T97},\cite{JH88}).

With the above preparation, we are now able to calculate the QCD vacuum tensor
susceptibility readily. Through the 2-point correlator of tensor current
$j_{\mu\nu}(x)=\bar{q}(x)\sigma_{\mu\nu}q(x)$,
\begin{equation}
  \Pi_{\mu\nu;\alpha\beta}(p)=\int{\rm d}^4xe^{ip\cdot x}\langle
0|T[j_{\mu\nu}(x)j_{\alpha\beta}(0)]|0\rangle, 
\end{equation}
the tensor susceptibility $\chi$ is defined as~\cite{HX96}
\begin{equation}
  \chi\equiv\frac{\Pi_{\chi}(0)}{6\langle\bar{q}q\rangle}, \hspace{1cm} 
  \Pi_{\chi}(q^2)\equiv\Pi_{\mu\nu;\mu\nu}(q^2).
\end{equation}
Using eq.~(\ref{Wick}), we get
\begin{eqnarray}
 & &\langle0|\bar{q}(x)\sigma_{\mu\nu}q(x)\bar{q}(0)\sigma_{\mu\nu}q(0)
             |0\rangle  \nonumber \\
 &=&{\rm tr}_{\gamma c}\int\frac{{\rm d}^4p}{(2\pi)^4}\sigma_{\mu\nu}
     \frac{-i{\not p}A(p^2)+B(p^2)}{X(p^2)}
     {\rm tr}_{\gamma c}\int\frac{{\rm d}^4q}{(2\pi)^4}\sigma_{\mu\nu}
     \frac{-i{\not q}A(q^2)+B(q^2)}{X(q^2)} \nonumber \\
 & &{}-\int\int\frac{{\rm d}^4p}{(2\pi)^4}\frac{{\rm d}^4q}{(2\pi)^4}
     e^{i(p-q)\cdot x}{\rm tr}_{\gamma c}\left [\sigma_{\mu\nu}
    \frac{-i{\not p}A(p^2)+B(p^2)}{X(p^2)}\sigma_{\mu\nu}
    \frac{-i{\not q}A(q^2)+B(q^2)}{X(q^2)}\right ]   \nonumber \\
 &=&-48N_c\int\int\frac{{\rm d}^4p}{(2\pi)^4}\frac{{\rm d}^4q}{(2\pi)^4}
     e^{i(p-q)\cdot x}\frac{B(p^2)}{X(p^2)}\frac{B(q^2)}{X(q^2)},
\end{eqnarray}
\begin{equation}
  \Pi_{\mu\nu;\mu\nu}(k)=-48N_c\int\int\frac{{\rm d}^4p}{(2\pi)^4}
  \frac{{\rm d}^4q}{(2\pi)^4}\frac{B(p^2)}{X(p^2)}
     \frac{B(q^2)}{X(q^2)}\delta(p-q+k),
\end{equation}
where $N_c=3$ is the number of colors, $X(s)=sA^2(s)+B^2(s)$. So we have 
the quantity
\begin{equation}
  {1\over 12}\Pi_{\chi}(0)=-\frac{3}{4\pi^2}\int^{\mu}_0{\rm d}ss
  \left [\frac{B(s)}{X(s)}\right ]^2,
\end{equation}
where $\mu$ is the renormalization scale which we chose to be 1 ${\rm GeV}^2$.

As a typical example, we let
$g^2D(s)=4\pi^2d\frac{\lambda^2}{s^2+\Delta}$, $d={12\over 27}$ and choose
three sets of different parameters for $\lambda$ and $\Delta$ by fixing the
pion decay constant in the chiral limit to $f_{\pi}=87$ MeV~\cite{T97}. In
Table~\ref{Tablechi} we display the values for ${\Pi_{\chi}(0)\over 12}$,
and the corresponding values for quark condensate $\langle\bar{q}q\rangle$ 
and the mixed quark-gluon condensate $g\langle\bar{q}\sigma Gq\rangle$ are
also displayed~\cite{T97}.
\begin{table}[ht]
\caption{\label{Tablechi}
The values of $\frac{\Pi_{\chi}(0)}{12}$ at $\mu=1$ ${\rm GeV}^2$ for
$g^2D(s)=(4\pi^2d)\frac{\lambda^2}{s^2+\Delta}$, $d={12\over 27}$ with three 
sets of different parameters. 
The quark condensate $\langle\bar{q}q\rangle$ and the mixed quark-gluon 
condensate $g\langle\bar{q}\sigma Gq\rangle$ are also presented.
} 
\begin{center}
\begin{tabular}{cc||c|c|c}
\hline\hline
$\Delta$ & $\lambda$ & $-\langle\bar{q}q\rangle^{1\over 3}$ &
$-g\langle\bar{q}\sigma Gq\rangle^{1\over 5}$ & $\Pi_{\chi}(0)/12$ \\ 
 $[{\rm GeV}^4]$ &[GeV]& [MeV] & [MeV] & [${\rm GeV}^2$]  \\ \hline\hline   
 $10^{-1}$       &1.77 & 183   & 460   &  -0.0016         \\ \hline 
 $10^{-2}$       &1.33 & 178   & 456   &  -0.0014         \\ \hline
 $10^{-4}$       &0.95 & 175   & 458   &  -0.0013         \\ \hline\hline   
\end{tabular}
\end{center}
\end{table}
The values of quantity ${\Pi_{\chi}(0)\over 12}$ vary with different input
parameters for a specific gluon 2-point function.

Our results
\begin{equation}
 \frac{\Pi_{\chi}(0)}{12}=-(\mbox{0.0013--0.0016})\ {\rm GeV}^2
\end{equation}
are much smaller than the estimations which were obtained recently~\cite{AS00}
from QCD sum rules with nonlocal condensates 
\begin{equation}
 \frac{\Pi_{\chi}(0)}{12}=-0.0055\pm 0.0008\ {\rm GeV}^2
\end{equation}
and from the standard sum rules
\begin{equation}
 \frac{\Pi_{\chi}(0)}{12}=-0.0053\pm 0.0021\ {\rm GeV}^2.
\end{equation}
Their results are similar to the estimations 
${\Pi_{\chi}(0)\over 12}=-0.008$ ${\rm GeV}^2$ given by Belyaev and Oganesian
from QCD sum rules~\cite{VA97} and 
${\Pi_{\chi}(0)\over 12}=-(\mbox{0.0083--0.0104})$ ${\rm GeV}^2$ from the 
chiral constituent quark model~\cite{WM98}. The earliest estimation obtained by
He and Ji~\cite{HX95,HX96} has opposite sign. The tensor susceptibility
given by Kisslinger from the QCD sum rules for three-point functions is large 
in magnitude and also has opposite sign.

In conclusion, the QCD vacuum tensor susceptibility can be calculated from GCM 
other than QCD sum rules. The value of tensor susceptibility is 
uniquely determined by the self-energy functions $A$ and $B$
for a given quark-quark interaction, which is chosen to reproduce $f_{\pi}$. 
The values of tensor susceptibility obtained in GCM are smaller than all 
the previous estimations. 

\noindent {\large \bf Acknowledgements}

One of us (Yang) would like to thank Dr. Weimin Sun for helpful discussions.

\end{document}